\newif\ifsubmode
\submodefalse

\newif\ifprintfig
\printfigtrue

\newif\ifemulate
\emulatetrue

\ifsubmode
\documentclass[12pt,preprint]{aastex}
\received{}
\accepted{}
\journalid{}
\articleid{}
\else
   \documentclass{emulateapj}
   \submitted{{\it To be submitted for publication in ApJ}}
\fi
\usepackage{multirow}

\newcommand{\Msun}{\mbox{\,$M_{\odot}$}}

\def\spose#1{\hbox to 0pt{#1\hss}}
\def\simlt{\mathrel{\spose{\lower 3pt\hbox{$\mathchar"218$}}
     \raise 2.0pt\hbox{$\mathchar"13C$}}}
\def\simgt{\mathrel{\spose{\lower 3pt\hbox{$\mathchar"218$}}
     \raise 2.0pt\hbox{$\mathchar"13E$}}}

\shorttitle{Dual Origin of Stellar Halos}
\shortauthors{Zolotov~et~al.}

\begin{document} 


\title{The Dual Origin of Stellar Halos}
\author{Adi\ Zolotov\altaffilmark{1,2}, Beth\ Willman\altaffilmark{2},
  Alyson\ M.\ Brooks\altaffilmark{3},  Fabio\
  Governato\altaffilmark{4},Chris\ B.\ Brook\altaffilmark{5}, David\
  W.\ Hogg\altaffilmark{1}, Tom\ Quinn\altaffilmark{4}, Greg\ Stinson\altaffilmark{6}}

\altaffiltext{1}{Center for Cosmology and Particle Physics, Department of Physics, New York University, 4 Washington Place, New York, NY 10003; az481@nyu.edu}
\altaffiltext{2}{Haverford College, Department of Astronomy, 370 Lancaster Avenue, Haverford, PA 19041; bwillman@haverford.edu}
\altaffiltext{3}{California Institute of Technology, M/C 350-17, Pasadena, CA 91125}
\altaffiltext{4}{Department of Astronomy,University of Washington, Box 351580, Seattle, WA 98195}
\altaffiltext{5}{Jeremiah Horrocks Institute, University of Central Lancashire, Preston, PR1 2HE, UK}
\altaffiltext{6}{Department of Physics and Astronomy, McMaster University, Hamilton, Ontario, L8S4M1, Canada}

\date{July 8, 2009}

\begin{abstract}

  We investigate the formation of the stellar halos of four simulated
  disk galaxies using high resolution, cosmological SPH + N-Body
  simulations. These simulations include a self-consistent treatment
  of all the major physical processes involved in galaxy
  formation. The simulated galaxies presented here each have a total
  mass of $\sim 10^{12} M_{\odot}$, but span a range of merger
  histories. These simulations allow us to study the competing
  importance of in-situ star formation (stars formed in the primary
  galaxy) and accretion of stars from subhalos in the building of
  stellar halos in a $\Lambda$CDM universe. All four simulated
  galaxies are surrounded by a stellar halo, whose inner regions
  (r$<20$ kpc) contain both accreted stars, and an in-situ stellar
  population. The outer regions of the galaxies' halos were assembled
  through pure accretion and disruption of satellites.  Most of the
  in-situ halo stars formed at high redshift out of smoothly accreted
  cold gas in the inner 1 kpc of the galaxies' potential wells,
  possibly as part of their primordial disks. These stars were
  displaced from their central locations into the halos through a
  succession of major mergers. We find that the two galaxies with
  recently quiescent merger histories have a higher fraction of
  in-situ stars ($\sim 20-50\%$) in their inner halos than the two
  galaxies with many recent mergers ($\sim 5-10\%$ in-situ
  fraction). Observational studies concentrating on stellar
  populations in the inner halo of the Milky Way will be the most
  affected by the presence of in-situ stars with halo kinematics, as
  we find that their existence in the inner few tens of kpc is a
  generic feature of galaxy formation.

\end{abstract}
\keywords{Galaxy: formation --- Galaxy: halos --- galaxies: formation -- galaxies: halos ---
          methods: N-Body Simulations}
\section{INTRODUCTION}\label{intro_sec}

Accreted stellar halos are a natural consequence of galaxy formation
in a $\Lambda$CDM universe, in which the smallest structures collapse
first and galaxy formation is thought to proceed bottom-up
\citep[e.g.][]{White1978,Searle1978}. Halos, composed of both dark and
baryonic matter, grow by merging with other halos. While the gas from
mergers and accretions loses its energy through cooling and settles
into a disk, the accreted stars and dark matter form a halo around the
galaxy. Galactic stellar halos are thus predicted to be built from
multiple accretion events starting from the first structures to
collapse in the Universe. 

It is for this reason that much work has focused on using
kinematically identified halo stars to look for fossil records of the
galaxy's accretion history. The possibility remains, however, that
stars formed within the potential well of a galaxy can become
displaced from the inner-most regions into a kinematic
stellar halo. In order to draw conclusions about the accretion history of a
galaxy based on halo stars, one must know how abundant such in-situ
stars are in the halo, and how they may dilute the signatures of
accretions and mergers.

 Owing to the low surface brightness of halos, the only presently well
 studied stellar halos are those of the Milky Way and M31. Current
 observational studies of the Milky Way's halo find strong evidence
 that its outer regions ($r \geq 20$kpc) were at least partially
 assembled through accretion events. Numerous stellar streams believed
 to have originated from small accreted galaxies have been observed in
 the Milky Way's outer halo \citep[e.g.][]{Newberg2002,
 Yanny2003,Belokurov2006, Grillmair2009}. Multiple wraps of stars
 believed to have been tidally removed from the Sagittarius dSph
 \citep{Ibata1994,Majewski2003} highlight the process of disrupting
 satellites in the Milky Way's potential, and serve as strong evidence
 of hierarchical merging. \citet{Bell2008} have found, using F and G
 turn off stars in SDSS, that the distribution of halo stars is highly
 structured, and the outer halo is consistent with being formed almost
 entirely from accreted subhalos. M31's stellar halo is observed to
 contain possibly even more substructure and streams than what is
 observed around the Milky Way \citep[e.g.][]{Ibata2001, Ibata2007,
 Gilbert2007}, suggesting its recent merger history may have been more
 active. Despite the plethora of streams observed around the Milky Way
 and M31, both the fraction of halo that is accreted, as well as the
 number of distinct accretion events leading to the observed
 substructure, is unclear.

 Although the nearby halo of our Galaxy has been studied more
 extensively than the outer halo, in some ways its properties are
 less well understood. This is due in part to the short dynamical
 timescales close to the disk, where accreted stars would no longer
 betray their origin through spatial structures. It is in these inner
 several tens of kpc of the halo that one might expect dissipative
 processes to contribute to the overall stellar
 population. Observational evidence of the contribution of accretion
 to the inner halo is slowly growing. \citet{Helmi1999} have detected
 two streams in the solar neighborhood by studying the angular
 momentum of stellar orbits. \citet{Morrison2009} have also recently
 shown that the angular momentum distribution of a sample of stars
 with a median height above the Galactic plane of 2 kpc does not seem
 to be smooth, possibly indicative of an accretion origin. A hybrid
 theory in which the Milky Way was assembled through a combination of
 hierarchical accretions and dissipative processes has been used by
 some to understand properties of field halo stars
 \citep[e.g.][]{Norris1994, Carollo2007}. \citet{Chiba2000}, using a
 large set of halo stars selected with no kinematic bias, find a
 vertical gradient in the mean rotation of metal poor stars, a
 signature of dissipational processes. \citet{Ibata2007}, using number
 count maps, have shown that a large fraction of the halo of M31
 appears to be rather smooth. The authors interpret these observations
 to suggest that some of the halo of M31 may have formed through
 dissipative processes. However, the degree to which dissipation is
 important in the formation of stellar halos, or one of their
 components, remains uncertain.
 
 Simulations have also been extensively used to investigate the
 formation of galactic halos.  The work of \citet{Bullock2005}
 combines dark matter only N-body simulations with semi-analytic
 models to study the build up of stellar halos via accretions alone.
 Their work has been highly successful in explaining the discrepant
 chemical abundances of Milky Way satellite galaxies and field halo
 stars \citep{Font2006a,Font2006b, Johnston2008}. While this powerful
 and efficient tool is particularly suited to better understand
 progenitors and signatures of accretion events, it can not tell us
 about the possible relative importance of dissipative processes in
 the assembly of a galactic halo. \citet{Brook2003a} have used N-Body
 + SPH to simulate the formation of disk galaxies in a self consistent
 way, following the evolution of dark matter, gas and star
 formation. They find that recently accreted satellites contribute
 stars to the halo which can be distinguished from an early formed
 halo through phase-space information. \citet{Abadi2006}, using a set
 of fully cosmological simulations, find that in-situ stars dominate
 the inner 20 kpc of their galaxies, while accreted stars make up the
 majority of the outer regions. While the Abadi work distinguishes
 between the accreted and in-situ populations in their galaxies, the
 work does not investigate the possible importance of in-situ stars in
 simulated stellar halos.

 In this paper, we use four high resolution cosmological SPH + N-Body
 simulations to investigate the origin of stellar halos. These
 self-consistent simulations permit the study of the competing
 importance of in-situ star formation and accretion from subhalos in
 the formation of stellar halos in a $\Lambda$CDM
 universe. Furthermore, the supernova feedback and cosmic UV
 background implemented in our simulated galaxies results in a
 luminous satellite population that is more similar to the Milky Way's
 than those in previous N-Body + SPH studies. This improvement is
 essential to model realistic stellar halos. We quantify the
 degree to which the halos are accreted, and note the effect of
 accretion histories on the overall formation of the stellar
 halos. While past numerical efforts have concentrated on the growth
 of stellar halos through pure accretion, we show that both accretion
 and in-situ star formation contribute to the inner regions of stellar
 halos.

 The paper is organized as follows. Section 2 describes the details of
 the simulations used, and the methods used to determine the origin of
 halo stars. In Section 3 we discuss the dual origin of the stellar
 halos of our simulated galaxies, and investigate how the properties
 of the in-situ and accreted populations in the stellar halos vary
 with merging history. Section 4 discusses possible numerical and
 resolution effects on the formation of the stellar halos in our
 simulations, as well as the effect of feedback on the accreted
 fraction. Section 5 discusses the connection to observational results
 and concludes.

\begin{deluxetable}{lcccccc}
\tabletypesize{\scriptsize}
\tablecaption{PROPERTIES OF SIMULATED GALAXIES}
\tablewidth{0pt}
\tablehead{
\colhead{Run} &
\colhead{$M_{vir}$}&
\colhead{$N_{tot}$ at z=0} &
\colhead{$M_{particle}^{DM}$} &
\colhead{$M_{particle}^{\ast}$} &
\colhead{$\epsilon$\tablenotemark{a}}\\
\colhead{} &
\colhead{$\Msun$} &
\colhead{dark+gas+stars}&
\colhead{$\Msun$} &
\colhead{$\Msun$} &
\colhead{kpc}
}
\startdata
MW1hr&$1.1 \times 10^{12}$& $ 4.9 \times 10^6$& $ 7.6 \times 10^5$&$ 2.7\times 10^4$&0.3\\
MW1med&$1.1 \times 10^{12}$&$ 6.1\times 10^5$&$6.1 \times 10^6$&$ 2.2\times 10^5$&0.3\\
Gal1&$3.3 \times 10^{12}$&$ 3.7 \times 10^6$&$ 2.6 \times 10^6$ & $ 9.2 \times 10^4$&0.3\\
h277&$7.4 \times 10^{11}$&$ 2.3 \times 10^6$&$ 1.2 \times 10^6$ &$ 4.6 \times 10^4$&0.35\\
h285&$7.7 \times 10^{11}$&$ 3.0 \times 10^6$&$ 1.2 \times 10^6$&$ 4.6 \times 10^4$&0.35\\
\enddata
\tablenotetext{a}{Gravitational softening length}

\end{deluxetable}

\section{SIMULATIONS}\label{sec_sims}

The high resolution simulations used in this study were run to z = 0,
  as part of an ongoing simulation project.  The two more massive
  galaxies (MW1hr, Gal1) are described in detail in \citet[][hereafter
  G07]{Governato2007}, but have since been rerun at eight times the
  mass resolution and twice the spatial resolution. The reader is
  referred to G07 for the full details of these simulations. The
  remaining two lower mass galaxies (H277, H285) are a continuation of
  these efforts. H277 is described in \citet{Brooks2008}. We briefly
  summarize the salient features of these simulations here.

The halos used in this study were selected, based on their mass and
  merger history, from low resolution, dark matter only simulations
  run with GASOLINE \citep{Wadsley2004}, using WMAP year 1 cosmology for
  MW1hr and Gal1, and WMAP year 3 cosmology for H277 and H285.  Each
  halo was then resimulated at higher resolution using the volume
  renormalization technique \citep{Katz1993}. This approach simulates
  only the region within a few virial radii at the highest resolution,
  while still simulating a large enough box to account for the effect
  the large scale tidal field has on the angular momentum of the halo.

  We concentrate a large part of this work on MW1hr, the highest
  resolution N-body + SPH simulation of a Milky Way mass galaxy run to
  $z=0$ published to date. We single this simulation out for extensive
  study as it has a lower resolution counterpart, run with the
  identical initial conditions, but 1/8 the mass resolution of MW1hr,
  as well as several runs completed with different treatments of
  feedback. The previous extensive study done on MW1hr (see G07) permit
  the numerical and resolution tests necessary for this work. MW1hr
  has 4.9 million particles (dark matter + gas + stars) within its
  virial radius, and a mass of $1.14 \times 10^{12} \Msun$. We define
  the redshift of last major merger as the time at which a secondary
  galaxy, with a mass no less than 1/3 of MW1hr's mass, first enters
  the virial radius of MW1hr. This occurs at a redshift of 4, although
  the merger is not complete until $z\sim 2.5$. The merger history of
  MW1hr is thus similar to that of the Milky Way, which is thought to have
  experienced its last major merger at z $\sim2.5$
  \citep{Hammer2007}. The mass of star particles in this simulated run
  is $\sim 2.7 \times 10^4 \Msun$, high enough to resolve an
  in-falling dwarf galaxy of stellar mass $10^9 \Msun$ with more than
  $10^4$ star particles. A physically motivated recipe was used to
  describe star formation and supernovae feedback \citep{Stinson2006},
  with a uniform UV background that turns on at z=9, mimicking cosmic
  reionization
\citep{Haardt1996}. The properties of all the simulated galaxies used
in this work are described in Table 1.

All dark matter halos and subhalos in the simulation are identified
  using AMIGA's Halo Finder \footnotemark
  \citep[AHF,][]{Knebe2001}. Each halo is identified using a minimum
  of 64 dark matter particle members, above which the halo mass
  function converges \citep[G07,][]{Reed2003}. The properties of each
  simulation particle are output at z=6, z=5, z=4, and every 320
  Myr after until z=0, for a total of 43 time steps output.
\footnotetext{AMIGA Halo Finder, available for download at http://www.aip.de/People/aknebe/AMIGA}

\subsection{Selection of Halo Stars}\label{ssec_comp}

Before we can study the mechanisms that built-up the stellar halos of
our simulated galaxies, it is first necessary to identify the star
particles which make up the galactic halos. In both our study, as well
as in observational work, it is important that the
stellar samples used to draw conclusions about halo properties and
formation are not contaminated by a population of disk stars. Unless
an observational sample only includes stars that reside several disk
scale heights above a galaxy's mid-plane, the use of kinematic and
metallicity information is necessary to obtain a clean sample of halo
stars.  Some observational works rely on samples of stars chosen based
on a metallicity cut, where the calculated [Fe/H] is several tenths of
dex below that of disk stars
\citep[for example][]{Beers2000,Chiba2000}. Other observational
studies have used the proper motions of stars to isolate those with
halo kinematics \citep{Smith2009}, while other works combine both
metallcities and kinematics \citep{Carollo2007,Morrison2009}.

We use the full simulation information to identify the most complete
and uncontaminated set of halo stars possible. We have chosen to
define our halo sample using the full 3 dimensional phase space
information available for all stars in the simulations. Rather than
implementing a specific spatial cut, this approach allows us to study
the detailed origin of all stars belonging to a halo. If we had
instead chosen to define the stellar halo of each simulated galaxy as
the population of stars with distances greater than $\sim 2$ scale
heights above the galactic disk, the trends we report in the following
sections would be unchanged. However, our results then would not have been
applicable to the many detailed observational studies of the Milky
Way's halo that are focused on stars within $d<5kpc$. In a future paper, we
will investigate the possible observational consequences of this
predicted origin.

A kinematic analysis of the star particles in each galaxy was done to
decompose them into disk, bulge, and halo kinematic components. To
decompose the disk from the spheroid, we first align the angular
momentum vector of the disk with the z-axis. This is done in order to
place the disk in the x-y plane. We then calculate the angular
momentum of each star in the x-y plane, $J_z$, as well as the momentum
of the co-rotating circular orbit with similar orbital energy,
$J_{circ}$. In this definition, stars with circular orbits in the
plane of the disk will have $J_z/J_{circ} \sim 1$. Disk stars are
selected with a cut of $J_z/J_{circ} \geq 0.8$, which is equivalent to
an eccentricity cut of $e < 0.2$. This condition matches the
eccentricities observed of disk stars in the Milky Way
\citep{Nordstrom2004}.

Once the disk population has been identified, we define a sphere
centered on each galaxy, with a radius equal to that of the
disk's. For stars within this inner sphere, the spheroid (halo $+$
bulge) of each galaxy is defined using a cut in $J_z/J_{circ}$ such
that the net rotation of the spheroidal population equals zero. The
spheroidal component of the galaxy is further decomposed into halo and
bulge because a break in the mass density profile is observed.  Stars
whose total energy is calculated to be low, and hence are tightly
bound to the galaxy, are classified as part of the bulge. All other
stars are placed in the halo category.  The energy cut between the
halo and bulge is set in order to make a two component fit of the mass
profile with bulge stars within the ``break'' and halo stars outside
the ``break''. Bulge contamination is not a concern because the energy
cut also ensures that stars which reside close to the center of the
galaxy, and are classified as belonging to the halo, have orbits which
take them far from a classical bulge.  Moreover, observational studies
that rely on kinematics to select halo stars would have associated our
halo stars with a halo, not bulge, component. We show in Section 3
that the stellar populations we associate with the halo extend out to
regions far beyond what would be considered a bulge.
Stars which reside beyond the inner sphere defined by the disk's
radius are all classified as halo stars, regardless of their
kinematics.  We have verified that the halo population has a
rotational velocity with a Gaussian distribution.

The $J_z/J_{circ}$ cuts described above leave a set of stars whose
kinematics do not make either the disk or spheroid cuts. These stars
are dominated by thick disk and pseudo bulge stars. We note that using
kinematic information for decomposing galactic components unavoidably
leads to some slight overlap in populations.

For the remainder of the paper, we use the term ``halo stars'' to
refer to stars identified using this kinematic definition. There is no
bias for our selected halo in-situ stars to be prograde in any
galaxy. This indicates that we have not mistakenly classified a
significant number of disk stars (in particular thick disk stars) as
halo stars in our analysis.

\subsection{Halo Star Origin}\label{ssec_origin}

We trace each star particle located within the main galaxy at $z=0$
  back in time. We identify at each time step the dark matter (DM)
  halo to which each particle belonged using AHF. Stars are considered
  bound to a DM halo if they are identified as being a member of that
  halo for two consecutive timesteps. We follow the particles'
  histories back to $z=6$, as well as follow the gas particles from
  which stars have formed. This procedure results in four different
  classifications for stars' origin: accreted, in-situ, ambiguous, and
  other. These classifications are described in the following
  sections.

\subsubsection{Accreted vs. Insitu}
We use these stellar histories to distinguish between stars that were
  accreted and those that formed ``in-situ''.  Accreted stars formed
  inside of halos other than the primary dark matter halo, but through
  accretion and merging have become unbound to their progenitor and
  now belong to the primary halo. In-situ stars formed from gas
  particles bound to the primary. These stars formed within the
  primary potential well and will be referred to as the ``in-situ''
  sample. In Section 3.1 we show that while these in-situ stars reside
  in the halo at z=0, they formed near the center of the galaxies'
  potential wells, as part of their primordial disks. The gas particle
  progenitors of these in-situ stars ended up in the primary halo from
  both smooth gas accretion, particularly in cold flows
  \citep{Brooks2008}, as well as from subhalos that underwent gas
  stripping as they merged with the primary. All galaxies are traced
  back far enough in redshift, to $z=6$, that we can accurately
  determine the star formation history of the vast majority ($>98 \%$)
  of halo stars. Approximately 1\% of the stars in these simulations
  formed before z=6, and as such we can not determine their formation
  origin.  We can thus accurately distinguish the accreted stars from
  the in-situ stars for the vast majority of particles in each
  simulated run.

\subsubsection{Ambiguous Origin Due to Time Resolution}
Our simulations also contain stars whose formation origin is unclear
 due to the limited number of time steps output for each
 simulation. For example, if a gas particle spawns a star particle at
 approximately the same time that it first becomes bound to the
 primary, it is uncertain whether that star formed in the primary or
 in its original subhalo. Such stars are referred to as ``ambiguous''
 in the reminder of the paper. These stars comprise less than $10\%$
 of all the kinematically identified halo stars in all of the
 simulations (see Table 2).

\subsubsection{``Other'' Stars in the Simulation}
To ensure that the in-situ stars are a robust feature of these
simulations, we now investigate whether this population of halo stars
were formed preferentially near the density and/or temperature
thresholds for star formation, where spurious star formation may occur
due to numerical effects. The spatial resolution of these simulations
is $\sim 300$ pc, much coarser than the scales on which star formation
occurs in real galaxies.  It is therefore necessary to use a
prescription for star formation that captures globally the observed
trends in galaxies.  The prescription used here for star formation
enforces a local gas density greater than 0.1 cm$^{-3}$ and a
temperature colder than $30,000$ K at which stars form.\footnotemark
\footnotetext{This temperature floor is due to our adopted cooling curve, 
which assumes primordial abundances without metals, does not yet
include molecular cooling, and thus cannot extrapolate below
$\sim$10,000 K.  It is assumed that dense gas at this temperature will
continue to rapidly cool.}  If a gas particle matches these criteria,
there is some probability that it will form a star particle.  This
probability is scaled to match the observed Kennicutt-Schmidt law for
star formation (see \citealp{Stinson2006} for further details).

We found that in-situ stars that formed at a look back time of 10 Gyrs
or more did not form near the density and temperature cutoffs. They
formed under the same conditions as stars with similar ages in the
disk of the galaxy.  However, in-situ halo stars in the simulation
younger than 10 Gyr old formed at preferentially lower densities and
higher temperatures than their young disk counterparts, near the
implemented cutoff for density threshold. While we have no reason to
believe these late forming in-situ stars are a numerical artifact,
changing our prescription for star formation would possibly affect
whether or not these in-situ stars had formed at all in the
simulation. \citet{Maller2004} have shown that accounting for hot halo
gas instabilities may be important in understanding the total baryonic
mass of a galaxy. These authors show that the fragmentation of the hot
halo gas results in hot low density gas which does not cool, and is
able to survive in the halo of a galaxy. This might explain the young
in-situ stars in our simulation that formed from a low density hot
gas. To be maximally conservative on our prediction of the extent to
which in-situ stars compose the halo, we have not included in-situ
stars that formed within the last 10 Gyr in any of the following
analysis. The broad conclusions of this paper are not affected by
whether or not we choose to ignore these intermediate and young
in-situ stars.  We refer to these stars as ``other'' for the reminder
of the paper, and only consider insitu stars formed more than 10
Gyr ago when discussing ``in-situ'' stars.  The total mass contributed
by other and in-situ stars to each simulated galaxy is listed in Table
2.

\section{DUAL ORIGIN FOR STELLAR HALOS: A GENERIC FEATURE OF GALAXIES}
\begin{figure*}[t!]
\epsscale{1.0}
\plottwo{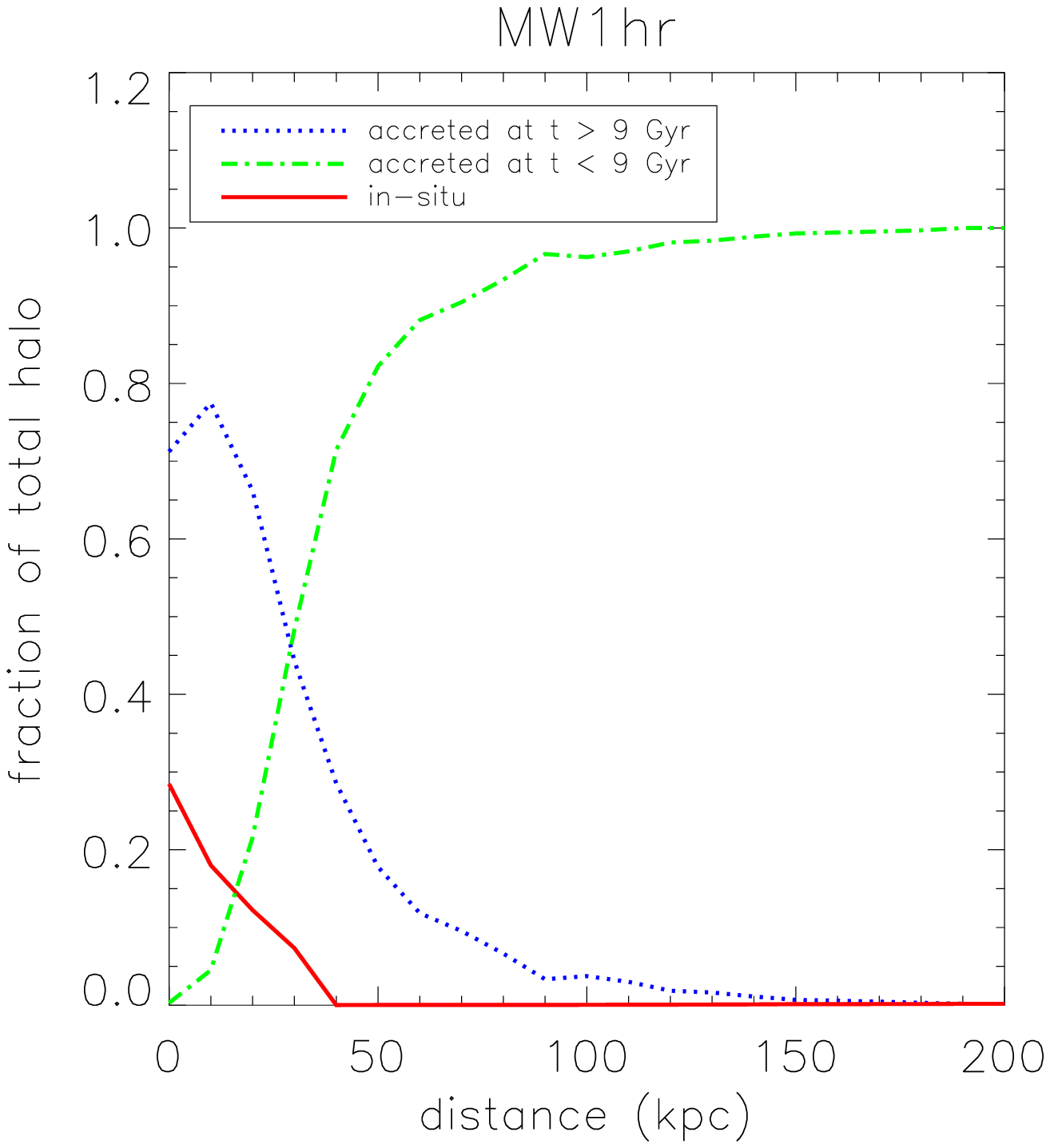}{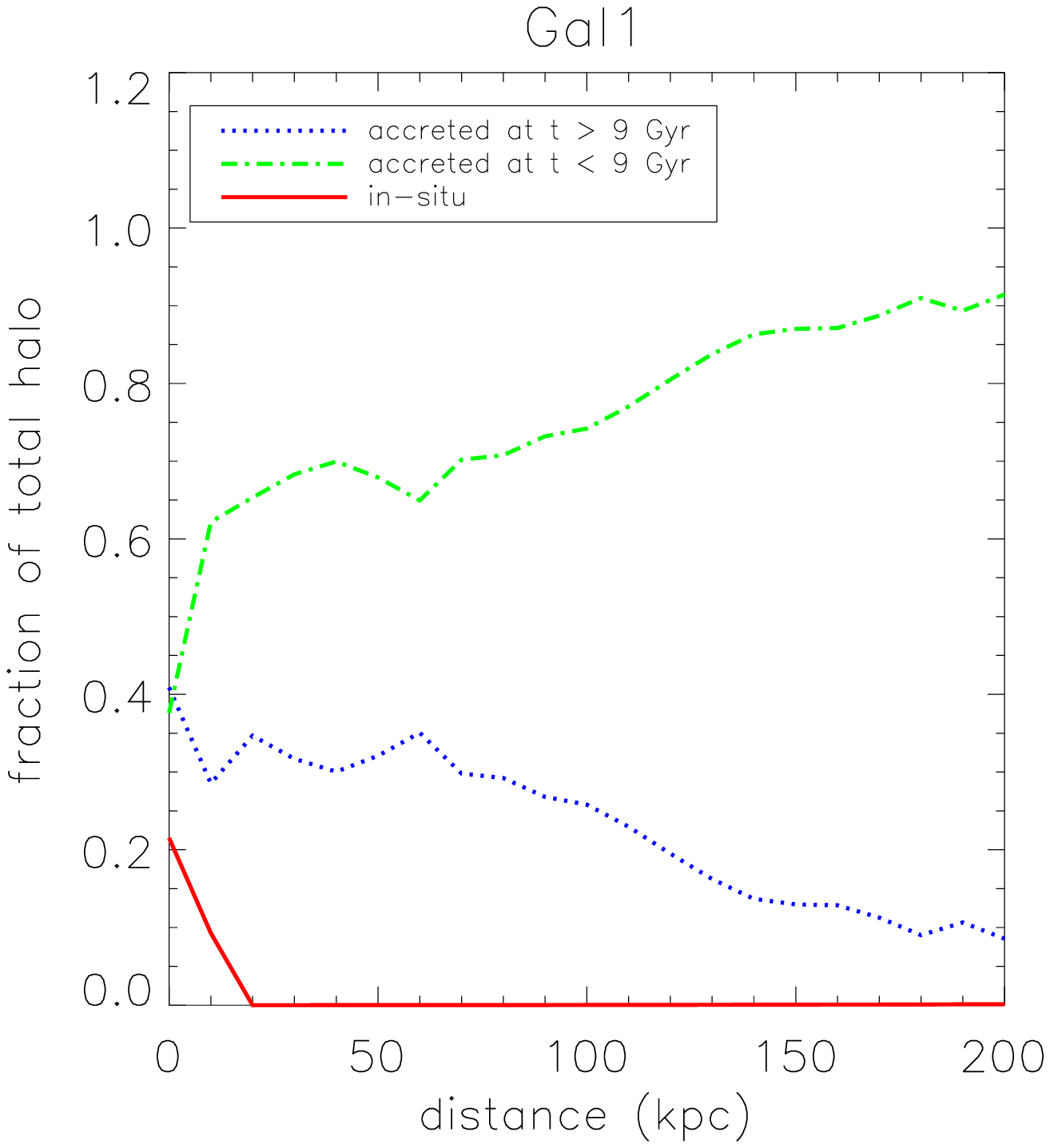}
\epsscale{1.0}
\plottwo{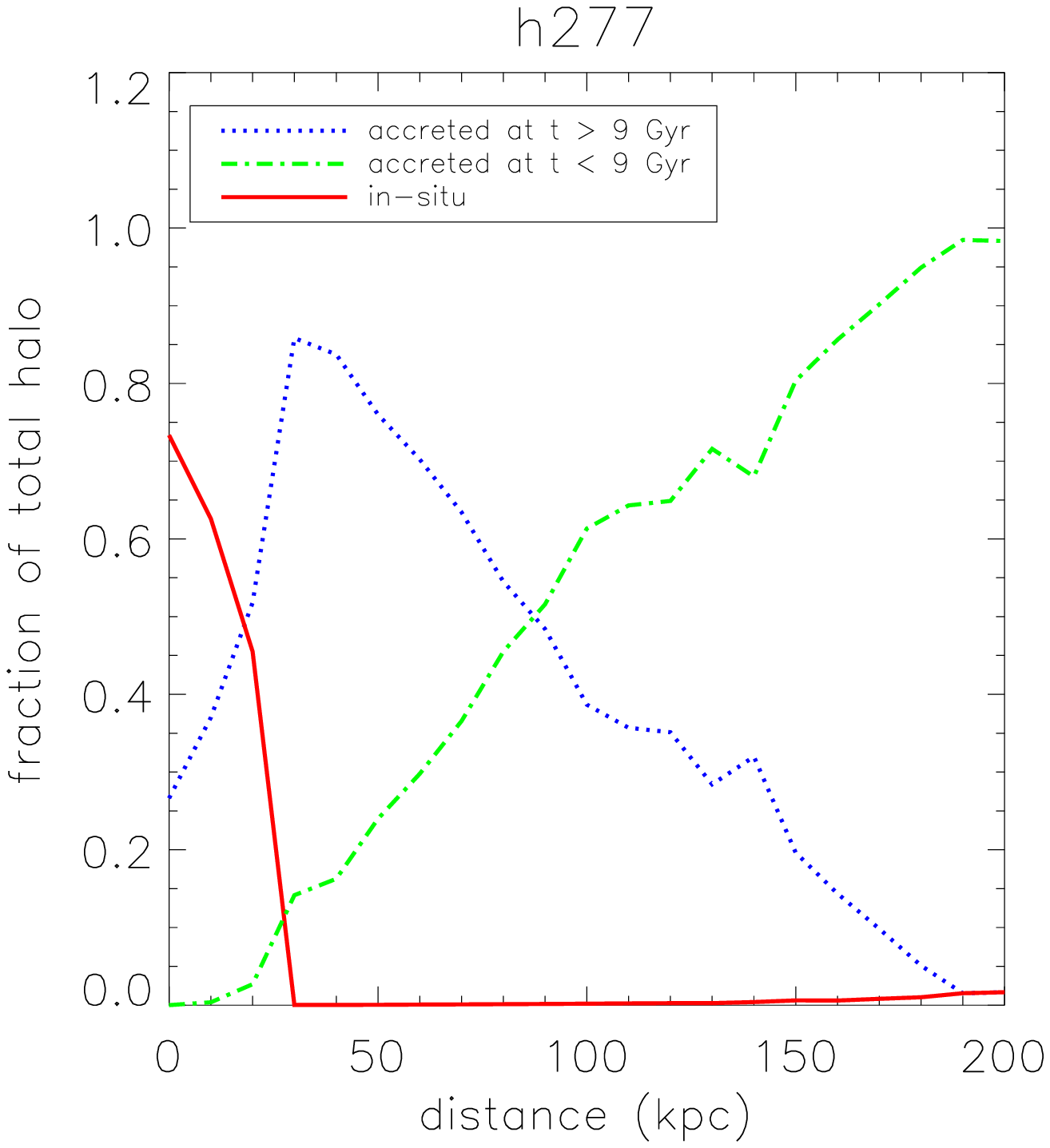}{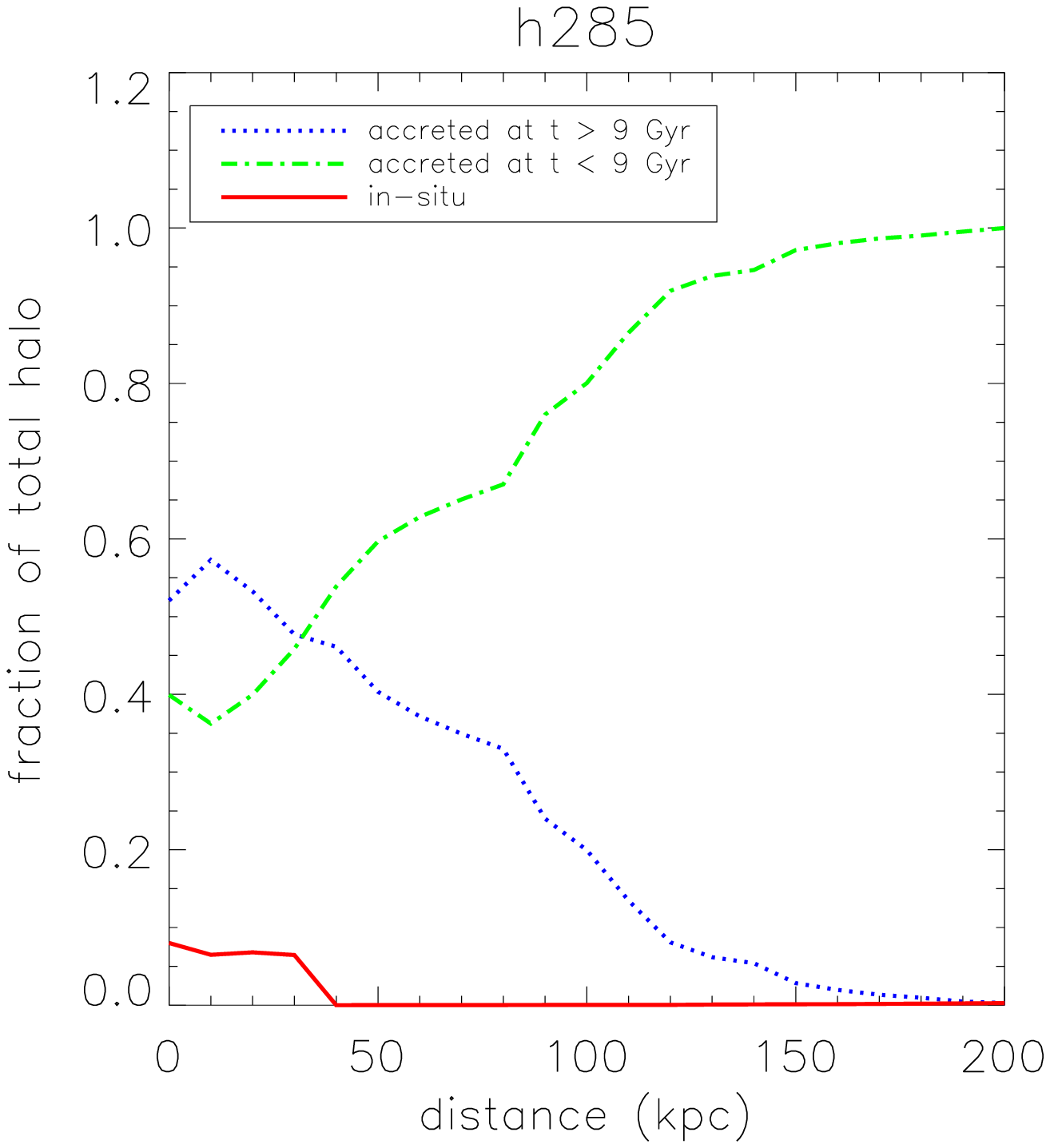}

\caption{
The relative contribution of accreted and in-situ stars to the stellar
halos of the simulated galaxies. Accreted stars are separated by the
time at which the stars first became bound to the primary halo. The
blue dotted line, and the green dash-dot line show the accreted stars
which became bound to the primary galaxy more than 9 Gyr ago, and less
than 9 Gyr ago, respectively, while the red solid line shows the
in-situ stars. Stars whose origins are unknown, as described in
Section 2.2, were omitted in this figure, and their fractional
contribution ignored.
\label{fig_radial}}
\end{figure*}

The presence of both in-situ and accreted stars is a generic feature
of the kinematically defined halos of all four simulated galaxies.
Each simulated stellar halo was found to be composed of a purely
accreted outer halo, and an inner halo where both accreted and in-situ
populations reside.  This is illustrated by
Figure~\ref{fig_radial}. The red solid lines in this figure show the
fractional contribution of in-situ stars to the stellar halos as a
function of distance. The blue dotted lines and green dash-dot lines
show the same for stars accreted to the halo at lookback times of more
and less than 9 Gyr, respectively. This figure shows that all four
stellar halos have an in-situ component that primarily resides within
their inner $\sim 20$ kpc, and that the accreted component of each
stellar halo extends to more than 100 kpc from the center of each
galaxy. The outermost regions of the stellar halos of all four
simulated galaxies are dominated by stars accreted within the last 9
Gyr.  The central location of all four in-situ populations shows that
their presence is most relevant for observational studies that are
limited to the inner halo of the Milky Way.  Table 2 lists the
fractional contribution, as well as the total mass, of the in-situ and
accreted populations in each of the four simulated halos in our
sample.

\subsection{In-Situ Star Formation}
We go on to study in detail the formation properties of in-situ stars
in the halo of MW1hr. As explained in Section 2, we focus on this
particular simulated halo because of the extensive numerical and
resolution testing which it has undergone. We find that the in-situ
stars of Gal1, h277, and h285 formed and populated their stellar halos
by the same qualitative process described below for MW1hr. 

In order to understand where the in-situ stars in the halo of MW1hr
originated, we studied the gas particles that formed this stellar
population. The vast majority of the gas progenitors were brought into
the primary galaxy in smooth gas flows. Fewer than $2\%$ of the in-situ
stars were formed from gas stripped from accreted subhalos. Cold gas
flow was found to also be the dominant contributor to the disk stars
of MW1hr
\citep{Brooks2008}.

To investigate how in-situ stars came to form part of the stellar halo
of MW1hr, we first study the locations at which these stars formed
relative to the center of the galaxy. Figure 2 shows the radial
distribution of in-situ stars at the time of their formation relative
to the center of the primary (black solid line). This shows that
in-situ stars form near the very center of MW1hr's potential well. 

Approximately $70\%$ of the in-situ stars were formed by z $\sim
3$. Between $2 < z < 3$, MW1hr experienced at least 3 significant
mergers with mass ratios $\frac{M_{MW1hr}}{M_{subhalo}} < 10$, defined
at the time of virial radius crossing. Substantial mergers disrupt the
orbits of stars in both the primary and merging galaxy. As two
galaxies of approximately equal mass collide, their potentials change
rapidly, causing the energies of their stars to change as well. The
violent relaxation process will cause some stars to gain energy, and
alter their initial orbits. We find that by z=2 (red dotted
line in Figure 2), the distribution of in-situ stars is no longer
sharply peaked at the inner 1 kpc of the galaxy, but is instead peaked
at 5 kpc from the galactic center. This transformation from centrally
located orbits to halo orbits was most likely due to the major mergers
which MW1hr experienced between $2<z<3$. After z=2, MW1hr does
not undergo any more mergers with such small mass ratios, and so we
expect that the radial distribution of in-situ stars will not
change. Indeed, we find that their distribution at the present day
(blue dashed line in Figure 2) is very similar to that at z=2.

The origin of the halo in-situ stars in cold gas flows,
and their formation in the innermost regions of MW1hr's potential well
all highlight that this population of stars formed in a rapid inflow
and collapse of gas that also formed the disk of MW1hr, but were
displaced through merger events into the halo. The process by which
these insitu stars formed is reflected in that their relative
importance in each stellar halo rises toward the center of each
simulated galaxy.

\begin{figure}[t!]
\epsscale{1.0}
\plotone{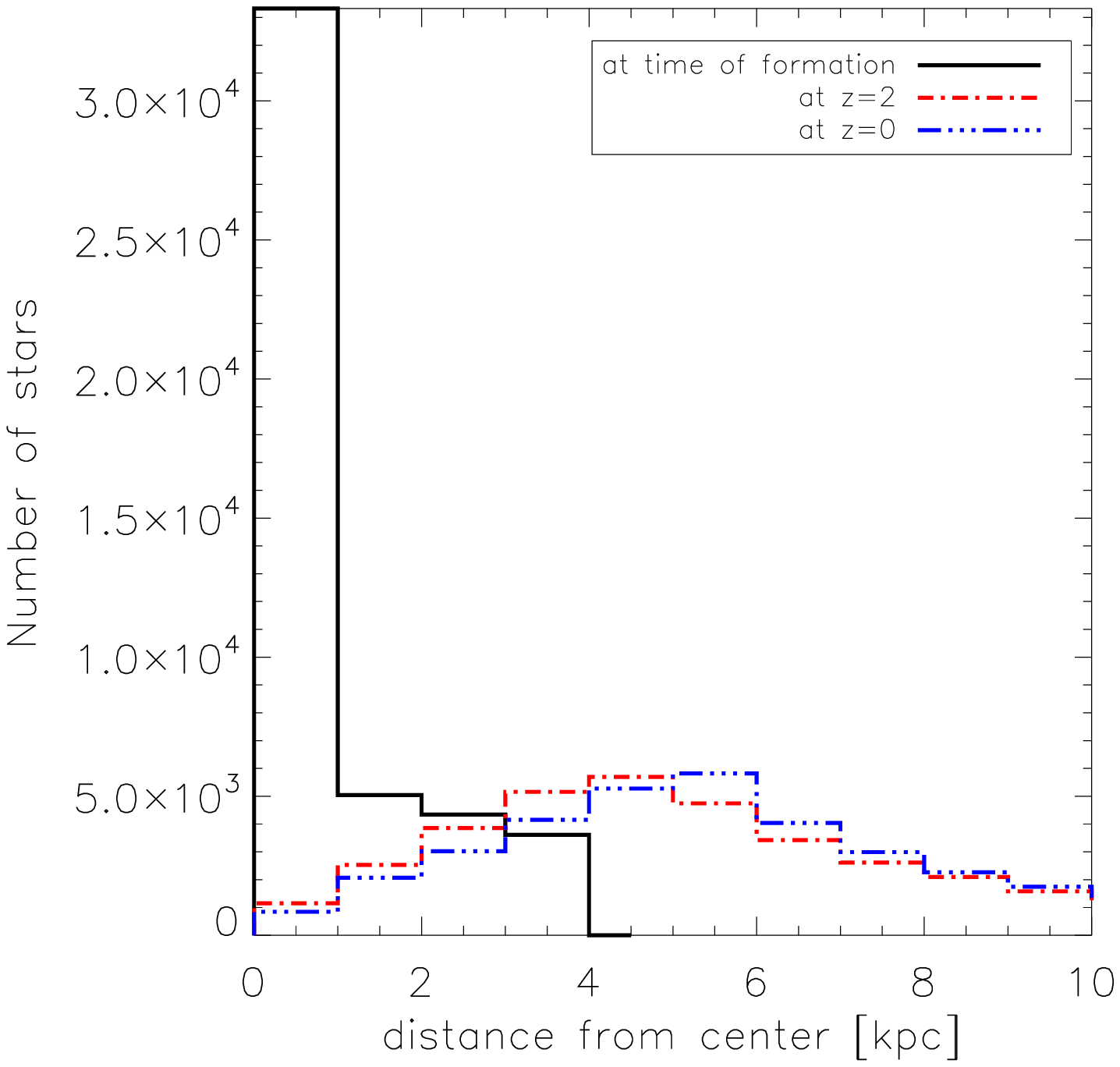}
\caption{The radial distribution of MW1hr's in-situ halo stars at the time of their formation (black solid line), at z=2 (red dotted line), and at present day (blue dashed line), relative to the center of MW1hr.
\label{fig_toomre}}
\end{figure}

\subsection{Comparing The Accreted Halos With Other Works}
Much work has been done recently with dark matter only N-body
simulations coupled with semi-analytic models to study the growth of
Milky Way-like galactic halos via pure accretion of stars. In this
section, we compare results from such numerical studies with our
findings for MW1hr's accreted halo. We focus here on MW1hr as it was
chosen to have a total mass, and merger history similar to that of the Milky
Way \citep{Governato2007}, making it a good candidate to compare with
previous works which have also focused on simulating Milky Way-like
stellar halos. We find that the accreted halo of MW1hr displays
properties which are typical in comparison with other numerical
studies.

As shown in the top left panel of Figure 1, accreted stars dominate
the halo of MW1hr at all radii.  While the innermost 20 kpc of the
halo contains both accreted and in-situ stars, the halo beyond this is
highly dominated by accretion. The earliest accreted stars, which
became bound to the primary halo of MW1hr at a lookback time of $\sim
9$ Gyr or more, dominate the accreted population of the halo out to 30
kpc from the center. In the halo's outer regions, however, the
majority of stars were accreted more recently.

The accretion history of MW1hr's stellar halo agrees well with the
numerical study of \citet{Helmi2003b}. These authors found that $\sim
90\%$ of stars are in place in the inner 10 kpc of a Milky Way-like
halo $10$ Gyr ago, while the mass growth of the outer regions is more
continuous. The work of \citet{Font2006a} also found that nearly
$80\%$ of the stars in the inner 20 kpc of simulated halos were in
place at a lookback time of 9 Gyr. \citet{Font2006b} have shown in
their numerical work that the inner regions of halos are typically
assembled from a few satellites with stellar masses $M_\star \sim 10^9
\Msun$. In MW1hr, the majority ($75\%$ by mass in stars) of the accreted 
mass in the inner 50 kpc originated in two subhalos, each with total
mass $\sim 2.5 \times 10^{10} M_{\odot}$.

The accreted stars which reside in the halos of Gal1 and H285 differ
significantly from those found in the stellar halo of MW1hr. Unlike
the accreted population of MW1hr, the majority of the accreted stars
in the halos of Gal1 and H285 became bound to their respective stellar
halo more recently, as shown in Figure 1. This is likely due to the
fact that both Gal1 and H285 have had a more active recent merger
history, as described in the following section. The accreted stellar
halo of H277 was assembled at similar times to that of MW1hr, although
the accreted population of this stellar halo contributes far less to
the overall mass of the halo in comparison to MW1hr.

\begin{deluxetable*}{lcccc|c}
\tabletypesize{\scriptsize}
\tablecaption{STELLAR HALO ORIGINS}
\tablewidth{0pt}
\tablehead{
\colhead{Run} &
\colhead{total halo:}&
\colhead{accreted: fraction (\%)} &
\colhead{in-situ: fraction (\%)} &
\colhead{ambiguous: fraction\tablenotemark{a}}& 
\colhead{``other'' stars:\tablenotemark{b}}\\
\colhead{}&
\colhead{mass ($M_{\odot}$) }&
\colhead{mass ($M_{\odot}$)}&
\colhead{mass ($M_{\odot}$) }&
\colhead{mass($M_{\odot}$)}&
\colhead{mass ($M_{\odot}$)}\\
}
\startdata
MW1hr & &70 & 21 & 9 & \\
& $6.22\times 10^9$ & $3.8 \times 10^9$ & $1.14 \times 10^9$ & $4.99 \times 10^8$ & $7.8 \times 10^8$\\ \hline
MW1med & &66 & 25 & 9 &  \\
& $5.24 \times 10 ^9$ &$3.23 \times 10^9$&$1.21 \times 10^9$ & $4.42 \times 10^8$ & $3.57\times 10 ^8$\\ \hline
MW1thermal\tablenotemark{c} && 85 & 12 & 3 & \\
&$9.74\times 10^9$&$7.96\times 10^9$&$1.19\times 10^9$&$2.17\times 10^8$&$3.6\times 10^8$\\ \hline
Gal1& &85 & 12 & 3 & \\
&$3.39 \times 10^{10}$ & $2.84\times 10^{10}$ & $4.17\times 10^9$ & $1.06\times 10^9$ & $ 4.04 \times 10^8$\\ \hline
H277& &32 & 58 & 10 &  \\
&$9.03\times 10^9$&$2.7 \times 10^9$&$4.85\times 10^9$&$7.89 \times 10^8$&$6.9\times 10^8$\\ \hline
H285&& 87 & 7 & 6 & \\
&$2.93\times 10^{10}$&$2.2\times 10^{10}$&$1.71\times 10^9$&$1.56\times 10^9$&$3.9\times 10^9$\\ \hline
\enddata
\tablenotetext{a}{The origin of these stars is ambiguous due to the finite time resolution of the simulation, as described in Section 2.2.2}
\tablenotetext{b}{These stars are excluded from the analysis of this paper, as discussed in Section 2.2.3}
\tablenotetext{c}{MW1thermal was run with the same mass resolution as MW1med, but with a different supernova feedback recipe. This is discussed in further detail in Section 4.1}

\end{deluxetable*} 

\subsection{Possible Role of Merger History}
Despite the universal presence of both in-situ and accreted halo stars
in all of the simulations, the relative contributions of these dual
components varies widely among the four stellar halos. The fractional
contributions of these stellar populations to the simulated stellar
halos, as well as their total masses, are summarized in Table 2. The
fraction of accreted halo stars ranges from $\sim 30 - 85\%$, while
the fraction of in-situ halo stars ranges from less than 10$\%$ to
more than 50$\%$. The high and low ends of these ranges do not sum to
$100\%$ because of the small presence of the ambigious stars. Although
the fractional contribution of in-situ stars covers a wide range, the
total mass of these stars in all four simulations is substantial. The
mass of in-situ stars in both the stellar halos of H277 and Gal1 is $\sim 4
\times 10^9 M_{\odot}$, and $\sim 1 \times 10^9 M_{\odot}$ for MW1hr and 
H285.

All four of these simulated galaxies have similar total masses,
however they span a range of merger histories.  Three of the galaxies,
MW1hr, H277, and H285 each host a stellar disk with a mass of $\sim 2 \times
10^{10} M_{\odot}$, while the mass of Gal1's stellar disk is $\sim 6
\times 10^{10} M_{\odot}$.  The total halo mass of H277 and MW1hr
is $6-9 \times 10^9 M_{\odot}$, and $\sim 3 \times 10^{10} M_{\odot}$
for Gal1 and H285. The bulge to disk ratio (B/D) of each simulated
galaxy was obtained using a two component Sersic $+$ exponential fit
to the one dimensional radial profile of the I band surface brightness
maps made with the SUNRISE software package \citep{Jonsson2006}. The
B/D for MW1hr, Gal1, H277, and H285 are 0.37, 0.66, 0.29, and 1.2,
respectively (Brooks et al, in prep).  What role might merger history
play in the range of accreted/in-situ fractions observed in these four
simulations?  There is not an obvious trend between epoch of last
major merger and either in-situ fraction or average assembly time of
the accreted component; Gal1 and MW1 had their last 3:1 merger at
$z\sim 4$ and H277 and H285 at $z \sim 2.5$. However, there appears to
be a possible trend with the galaxies' recent merger histories.
Figure~\ref{fig_radial} illustrates one manifestation of this. The
accreted stellar halo components of MW1 and H277, the two galaxies
with the highest in-situ fraction, were largely assembled more than 9
Gyr ago, with the innermost regions of these halos containing very few
stars accreted in the last 9 Gyr. However, Gal1 and H285, which
contain few in-situ halo stars, have accreted stellar halos that, on
average, were assembled much more recently at all radii, with at least
$30\%$ accreted within the last 9 Gyr. It is important to recall that
we have defined accretion time as the time at which a star becomes
unbound from its subhalo, and bound to the primary dark matter halo,
which is a different, and invariably later, time than when the subhalo
entered the primary's virial radius.

In order to more explicitly study the merger histories of the
simulated galaxies, we traced the total number of luminous satellites 
that enter the virial radius at different
redshifts. In particular, we are interested in those satellites whose
masses are in the range $10< M_{primary}/M_{satellite} < 200$ at the
time of merging, as these are expected to be the building blocks of
the accreted component of stellar halos
\citep{Bullock2005}. Table 3 shows the number of such merged satellites
at redshifts smaller than $z=2$. The stellar halos of Gal1 and H285,
the two galaxies with more recently assembled stellar halos, have
continued to accrete a large number of satellites after z=2. The masses
of these galaxies' accreted stellar halos are comparatively large ($>
10^{10} M_{\odot}$), overwhelming the population of halo in-situ
stars. In contrast, MW1hr and H277, whose stellar halos were found to have
assembled at more ancient times, have had a more quiet recent merger
history. These galaxies merged with fewer satellites in the studied mass
and redshift range, and have the less massive accreted halos. MW1hr
and H277 also both have halos which host a comparatively larger in-situ population than
Gal1 and H285.

\begin{deluxetable*}{lccccc}
\tabletypesize{\scriptsize}
\tablecaption{LOW z MERGERS IN SIMULATED GALAXIES}
\tablewidth{0pt}
\tablehead{
\colhead{Run} &
\colhead{mass ratio}&
\colhead{mass ratio} \\
\colhead{}&
\colhead{[10:50]}&
\colhead{[50:200]}\\
}
\startdata
MW1hr& 2 & 5\\
Gal1& 2 & 10 \\
H277& 1 & 2\\
H285& 7 & 8\\
\enddata
\tablecomments{The number of merged satellites within the given mass ratio for each simulated galaxy ($M_{primary}/M_{satellite}$), at times more recent than a redshift of 2.}
\end{deluxetable*}

\section{Numerical and Resolution Studies}

Owing to a dual origin, the detailed properties of a stellar halo are
  a function of both i) the properties of the primary galaxy itself,
  and ii) the properties of the protogalaxies that merged to form its
  accreted component. Accurate modeling of a galaxy's stellar halo
  thus requires an accurate treatment of the formation and evolution
  of that galaxy over a wide dynamical range. In this section we
  discuss the effects of feedback, as well as numerical and resolution
  effects, on the overall formation of stellar halos in our simulated
  galaxies.

\subsection{The Effect of Feedback on the Simulated Galaxies}

  The four simulated galaxies used in this study have sizes, ages, and
  metallicities that match those observed at low redshift.
  G07 and \citet{Governato2008} showed that disk
  galaxies naturally form in a cosmological context in simulations
  such as those studied here, and have sizes that place them on the
  I-band Tully--Fisher relation (TF,
  e.g. \citealp{Giovanelli1997,Mcgaugh2005,
  Geha2006}). \citet{Brooks2007} used a set of GASOLINE simulations
  (including MW1) to show that galaxies with total $z =$ 0 halo masses
  in the range $3.4 \times 10^9 M_{\odot} < M_{\star} < 1.1 \times
  10^{12} M_{\odot}$ have (O/H) abundances that match both those
  measured by \citet{Tremonti2004} for more than 53,000 SDSS galaxies
  locally, and those measured by Erb et al.\,(2006) at $z =$ 2.
  \citet{Maiolino2008} demonstrated that galaxies simulated with
  GASOLINE (including the progenitors of H277 and H285) are also a
  good match to the observed mass--metallicity relationship at $z =$
  3.

The success of these simulations in matching the observed
mass--metallicity relationship at varying redshifts \citep{Brooks2007,
Maiolino2008} demonstrates that these simulations have made
substantial progress in overcoming the historic ``overcooling
problem'' \citep[e.g.][]{Mayer2008}. Gas in simulations has
traditionally cooled rapidly, forming stars quickly and early, and
thus producing a mass--metallicity relation that is too enriched at a
given stellar mass, particularly at high $z$.  In the simulations
examined here, supernova energy is deposited into the ISM mimicking
the blast-wave phase of a supernova (following the Sedov-Taylor
solution, details in
\citealp{Stinson2006}).  Cooling is turned off for gas particles within the
supernova blastwave radius for a period of time after the supernova event.
Supernova feedback regulates star formation efficiency as a function
of halo mass, resulting in the mass--metallicity relationship
described above.  This regulation of star formation also leads to
realistic trends in gas fractions, with our lowest mass galaxies being
the most gas rich \citep{Brooks2007}, and reproducing the observed
incidence rate and column densities of Damped Lyman $\alpha$ systems
at $z =$ 3 \citep{Pontzen2008}.

Feedback, with its role in regulating star formation, will also effect
the luminosity function of galaxies in a simulation, which will in
turn effect the properties of the accreted stellar halo.  For example,
if the simulated dwarfs have too many stars relative to dwarf galaxies
in the real universe, then we may overestimate the relative importance
of accreted versus in-situ stars in the halos.  We therefore compare
the luminosity function of satellites within the virial radius of our
two medium resolution galaxies, MW1med and MW1thermal.  We use these
two simulated runs because they have the same mass resolution, but
different implemented feedback mechanisms.  MW1med adopts the
blastwave model discussed above, and used for all of the high
resolution runs studied here.  In MW1thermal, supernova energy was
deposited as thermal energy to the surrounding gas particles but
cooling is not turned off in these particles.  These gas particles
will then quickly radiate away the deposited energy, and this
simulation will suffer from the overcooling problem.

Using the age and metallicity of each stellar particle in a satellite,
  we calculated the absolute V band magnitudes of the z = 0 satellites
  of MW1med and MW1thermal using the stellar population models of
  Starburst99 \citep{Leitherer1999, Vazquez2005}. Figure 3 shows the
  luminosity function of the satellites in MW1med and MW1thermal, as
  well as the observed luminosity function of Milky Way dwarfs, from
  \citet{Grebel2003}.  The satellites of MW1med are similar in total
  number and luminosity to the Milky Way's, except at the bright end
  of the luminosity function. The satellites of MW1thermal, however,
  are too numerous at all luminosities. The discrepancy between the
  luminosity function of MW1med and MW1thermal is due to the
  overcooling problem typical of simulations with thermal feedback,
  which causes satellites to form which are highly centrally
  concentrated and contain too many stars.  Because the MW1thermal run
  produces over-luminous dwarf galaxies, we expect that its halo will
  be more dominated by accreted stars than the halo of MW1med.  As
  shown by Table 2, which lists the accreted and in-situ fractions of
  both of these galaxies, the halo of the MW1thermal has $\sim 20\%$ more
  accreted stars in its halo relative to in-situ stars than MW1med
  ($85\%$ vs $66\%$) and $<$ 1/2 of the fraction of in-situ stars
  ($12\%$ vs $25\%$). While the total mass of in-situ halo stars is
  about the same for both simulated runs ($1.2 \times 10^9 M_{\odot}$), the
  total mass of accreted stars in the thermal run is $8 \times 10^9
  M_{\odot}$, and only $3.2 \times 10^9 M_{\odot}$ in the run with
  blast-wave feedback.

In summary, several features of the simulations we analyze here
contribute to bringing the number of luminous satellite galaxies into
good agreement with those observed around the MW:  Our uniform UV
background unbinds baryons from simulated halos with total masses
below $\sim 10^9 M_{\odot}$, creating ``dark'' halos and bringing down
the number of luminous satellite galaxies.  Our supernova feedback
scheme helps to significantly reduce the number of stars produced in
the remaining low mass satellite galaxies.  However, because our
satellites are still slightly too bright (ie, have too many stars),
our accreted fractions are an upper limit.  This only serves to
increase the potential relative contribution of in-situ stars in the
inner halos of galaxies.

\begin{figure}[t!]
\epsscale{1.0}
\plotone{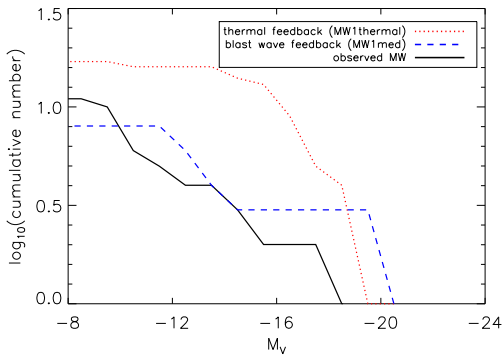}
\caption{The luminosity function of satellites for two simulated runs with different feedback prescriptions, as well as the observed luminosity function of the Milky Way's dwarfs, from \citet{Grebel2003}.
\label{fig_fract_noprog}}
\end{figure}

\subsection{Resolution Effects on Star Formation}

\citet{Brooks2007} found that several thousand particles were 
required inside of the virial radius of a halo in order for the star
formation history to converge.  For the high resolution simulations
studied here, this convergence criterion is met in halos with total
masses above a few $\times 10^9 M_{\odot}$. For the medium resolution
runs, the convergence limit is met in halos with total masses above $2
\times 10^{10} M_{\odot}$.  Resolution testing has shown that poorly
resolved halos have smaller total stellar masses than halos which are
fully resolved.  We therefore underestimate the contribution to the
primary stellar halos from dwarf galaxies with total mass below this
convergence limit. However, we find that dwarf galaxies less massive
than $10^{10} M_{\odot}$ contribute less than 5\% to the overall mass
in accreted stars in the high resolution stellar halos. Due to the
resolution effects discussed above, this small fraction of accreted
satellites will contribute even fewer stars at lower resolution than
they would if they were fully resolved, decreasing their overall
contribution to the accreted stellar halo of the medium resolution
runs (MW1med and MW1thermal). This effect is minimal due to the tiny
contribution of these halos. Our findings agree with the work of
\citet{Bullock2005}, who showed that satellites with total mass
greater than $2\times 10^{10} M_{\odot}$ contribute $75-90 \%$ of the
mass to the primary stellar halo.

Because the in-situ stars are forming deep in the potential well of
the high resolution main halo, we do not expect their formation
properties to vary with resolution because the star formation history
of the main halo is fully resolved.  In order to test this assumption,
we compared the in-situ formation in the high resolution MW1hr run
with the lower resolution MW1med run. We expect that if the formation
of these stars was caused by a numerical resolution effect, then their
fractional contribution to the simulated halos would change as the
resolution at which the galaxy is simulated changed. We find, however,
that at both resolutions, in-situ stars make up a substantial fraction
of the stellar halo of the Milky Way massed galaxies. Table 2 shows
that the overall in-situ fraction of the stellar halos of MW1hr and
MW1med are $21\%$ and $25\%$, repectively. The $4\%$ larger
contribution of in-situ stars to the stellar halo of MW1med is likely
due to the slight underestimating of accreted stars in this medium
resolution run, as described above.

\section{DISCUSSION \& CONCLUSIONS}
In this paper we have analyzed four SPH + N-Body simulations of
approximately $L^{\star}$ disk galaxies to investigate the formation
of their stellar halos. We have shown that the inner halos of all four
galaxies contain both stars accreted from merged satellites, as well
as in-situ stars formed within the primary galaxy. While these results
are theoretical, they are backed by recent observational studies of
the stellar halos of the Milky Way and M31.

Observational work has shown that the properties of Milky Way halo
stars exhibit signatures of a possible double
population. \citet{Carollo2007} have found that a sample of local halo
stars in SDSS can be thought of as two distinct, but overlapping,
components, with different spatial distributions, orbits and
metallicities. They attribute, as others have, these different stellar
properties to different formation mechanisms for the two halo
components. They conclude that the inner halo may have formed
dissipatively from the gas rich merger of early sub-galactic halos,
while the outer halo stars were accreted from subhalos which merged
with the Milky Way's dark matter halo. 

Observations of M31 have revealed that this giant spiral galaxy is
also surrounded by a large extended stellar population.  M31 has been
shown to have a distinct stellar halo with a surface brightness
profile similar to the Milky Way's, and a metallicity gradient with
the inner regions of its halo (R$ <30$ kpc) more metal rich than the
outer regions \citep{Guhathakurta2005,Kalirai2006}.
Using number count maps, \citet{Ibata2007} find a large
metal poor smooth component to the halo of M31, with a more metal
rich inner halo. These works highlight the possible signatures that
M31's stellar halo may, like the Milky Way, have two underlying
components, with different formation histories.

The results of such observational findings for the halos of the Milky
Way and M31, however, have previously lacked counterparts in
simulations as most of the numerical studies done to date on stellar
halos have concentrated on their build up through pure accretions. Our
work begins to bridge the gap by studying self-consistent simulations
where the contribution of in-situ stars to the overall halo have been
taken into account. Our main results are:

\begin{itemize}

{\item In-situ stars are a generic feature of kinematically defined
 stellar halos. These stars formed deep in the potential well of the
 primary galaxy and were later displaced to halo orbits as a result of
 mergers.}

{\item In-situ stars reside primarily in the inner few tens of kpc
of the galaxies' halos.}

{\item Stars accreted off satellites dominate the stellar halo at all
radii for three of the four galaxies studied. The accreted stellar halo
components were found to extend to out to a few hundred kpc.}

{\item The fractional contribution of in-situ stars to the
simulated galaxies' halos varies greatly, from $5 -50\%$.}

{\item The two galaxies in our sample whose halos had the smallest
contributions from in-situ stars were also those with accreted halos
that assembled less than 9 Gyr ago. This is likely because such
galaxies have more active recent accretion histories, which lead to
massive accreted halos. The large contribution of recently accreted
stars overwhelm the underlying in-situ populations in these halos. The
other two galaxies in our sample, with large contributions from
in-situ stars, had a significant fraction of their accreted stars in
place at times earlier than 9 Gyr ago. The relative importance of
in-situ stars in these halos is due to the recent quiet merger history of
these galaxies, which have led to less massive accreted halos. }

{\item Different supernova feedbacks greatly impact the formation of
simulated stellar halos. This underscores the necessity of carefully
implementing feedback to produce a realistic satellite luminosity
function, and to hence conduct realistic studies of halo formation.}

{\item The overall importance of in-situ stars to the stellar
halos found in this study is likely a lower limit. This is because i)
we have ignored a fraction of in-situ stars for numerical reasons, and
ii) the satellites of the simulated galaxies studied here are brighter than
those observed in the Milky Way. The calculated contributions of accreted stars
to the stellar halos are thus an upper-limit.}

{\item The qualitative trends of our results are unchanged if we had
instead used a spatially defined set of stellar halos.}

\end{itemize}

Our results suggest that galaxies with more active recent
mergers, like M31, may host halos where the relative importance of an
in-situ population is low. Galaxies, such as the Milky Way, with
little evidence for recent merger events, on the other hand, may have
a larger relative contribution from an in-situ population in its inner regions,
along with its accreted stars. However, a larger set of simulations
with a wide range in merger histories is needed to fully understand
the complex relationship between halo formation and mergers.

Vast observational resources are currently dedicated to the study of
the Milky Way's and M31's halos. Spatial, kinematic and chemical
abundance information of halo stars being collected and used to
attempt to reconstruct the accretion and formation history of the
galaxies. Our results highlight that such data sets, particularly ones
focused on local, and hence inner, halo stars, will contain a
significant population of stars born in-situ in the galaxies. These
in-situ stars may dilute the signatures of accretion history in
observations. Given the short dynamical timescales in the inner halo,
one may not expect to be able to discriminate these two populations
using their kinematics alone. However, given the very different sites
of their formation, the chemical abundances of in-situ and accreted
stars may be quite distinct. In a future paper, we will investigate
possible ways to observationally disentangle the two halo populations.

\section*{Acknowledgments}
We thank the anonymous referee for suggestions which helped to
extensively clarify the paper.  AZ thanks the Institute of Theory
\& Computation at Harvard-Smithsonian's Center for Astrophysics for
its hospitality during part of this work. BW and AZ thank the
Smithsonian Astrophysical Observatory and the Clay Fellowship for
partial financial support. We thank Joe Cammisa at Haverford for
computing support.  AZ was partially supported by New York
University's Horizon Fellowship. AB acknowledges support from the
Sherman Fairchild Postdoctoral Program in Theoretical
Astrophysics. DWH was partially supported by NASA grant
NNX08AJ48G. Simulations were run at ARSC, NASA AMES and Texas
Supercomputing Center. FG acknowledges support from a Theodore Dunham
grant, HST GO-1125, NSF ITR grant PHY-0205413 (also supporting TQ),
NSF grant AST-0607819 and NASA ATP NNX08AG84G. CBB acknowledges the
support of the UK's Science \& Technology Facilities Council
(ST/F002432/1).

\end{document}